\documentclass[article,12pt]{article}
\pdfoutput=1

\usepackage{jheppub}

\usepackage{amssymb}
\usepackage{amsbsy}
\usepackage{amsfonts}
\usepackage{amsmath}
\usepackage{epsfig}
\usepackage{cancel}
\usepackage{color}
\usepackage{subfigure}
\usepackage{graphicx}
\usepackage{epstopdf}
\usepackage{float}

\def\bea{\begin{eqnarray}}
\def\eea{\end{eqnarray}}
\def\be{\begin{equation}}
\def\ee{\end{equation}}
\def\beqn{\begin{eqnarray}}
\def\eeqn{\end{eqnarray}}
\def\beq{\begin{equation}}
\def\eeq{\end{equation}}

\def\Dslash{\not{\hbox{\kern-4pt $D$}}}
\def\pslash{\not{\hbox{\kern-4pt $p$}}}

\newcommand{\nn}{\nonumber}

\title{Dirac gauginos, R symmetry and the 125 GeV Higgs}

\author[a, b]{Enrico~Bertuzzo}
\author[c]{Claudia~Frugiuele}
\author[d]{Thomas~Gr\'egoire}
\author[e]{Eduardo~Pont\'on}

\emailAdd{ebertuzzo@ifae.es}
\emailAdd{claudiaf@fnal.gov}
\emailAdd{gregoire@physics.carleton.ca}
\emailAdd{eponton@ift.unesp.br}

\affiliation[a]{Institut de Physique Th\'eorique, CEA-Saclay, 91191 Gif-sur-Yvette, France}
\affiliation[b]{IFAE, Universitat Aut\`onoma de Barcelona, 08193 Bellaterra, Barcelona, Spain}
\affiliation[c]{Fermilab, P.O. Box 500, Batavia, IL 60510, USA}
\affiliation[d]{Ottawa-Carleton Institute for Physics, Department of Physics, Carleton University 1125 Colonel By Drive, Ottawa, K1S 5B6 Canada}
\affiliation[e]{ICTP South American Institute for Fundamental Research, and \\ Instituto de F\'isica Te\'orica - Universidade Estadual Paulista (UNESP), \\ Rua Dr. Bento Teobaldo Ferraz 271, 01140-070 S\~ao Paulo, SP Brazil}
\abstract{ 

We study a supersymmetric scenario with a quasi exact R-symmetry in
light of the discovery of a Higgs resonance with a mass of 125 GeV. In
such a framework, the additional adjoint superfields, needed to give
Dirac masses to the gauginos, contribute both to the Higgs mass and to
electroweak precision observables.  We analyze the interplay between
the two aspects, finding regions in parameter space in which the
contributions to the precision observables are under control and a 125
GeV Higgs boson can be accommodated.  We estimate the fine-tuning of
the model finding regions of the parameter space still unexplored by
the LHC with a fine-tuning considerably improved with respect to the
minimal supersymmetric scenario.  In particular, sizable
non-holomorphic (non-supersoft) adjoints masses are required to reduce
the fine-tuning.  }

\begin{document}

\maketitle

\section{Introduction}

The discovery of a 125 GeV particle closely resembling the Standard
Model (SM) Higgs~\cite{Chatrchyan:2012ufa, Aad:2012tfa} may represent
a challenge for Supersymmetry (SUSY).  Indeed, at least in its minimal
version, large loop contributions are needed to raise the mass of the
lightest Higgs boson to the observed value, the most relevant ones
coming from the stop system.  This points toward very heavy stops,
and/or large left-right stop mixing.

While this is perfectly consistent with the non observation of any
superpartner at the LHC, it is widely believed to be at odds with the
concept of naturalness, which requires light stops with small
left-right mixing.  Needless to say, after the first LHC run and the
Higgs discovery, understanding whether the concept of naturalness as
it stands is or not a principle followed by nature has become of the
utmost importance.

If we insist on naturalness, we need to consider alternatives to the
Minimal Supersymmetric Standard Model (MSSM).  An interesting
possibility is given by models with Dirac gauginos, which have relaxed
naturalness bounds on the gluino mass.  This is most welcome, since
being the gluino the most constrained particle after the first LHC
run, a relaxed naturalness bound on its mass gives less tension with
data.  The mechanism behind the improved naturalness is the generation
of Dirac gaugino masses through supersoft operators, which give only
finite contributions to scalar masses~\cite{Fox:2002bu}.  Models with
Dirac gauginos are also interesting from a purely phenomenological
point of view: first of all, squark pair production is suppressed at
the LHC due to the absence of Majorana mass insertions~\cite{Kribs:2012gx}.
Moreover, Dirac gaugino masses are compatible with the presence of a
global $U(1)_R$ symmetry, which would be otherwise broken by the
Majorana mass term.  The R-symmetry can be used as an alternative to 
R-parity to forbid operators leading to proton
decay~\cite{Hall:1990dga,Hall:1990hq}, but has far richer
consequences.  Indeed, the absence of $A$ terms, the $\mu$ term and
Majorana gaugino masses has a drastic beneficial effect on the SUSY
flavor problem~\cite{Kribs:2007ac}.

A peculiar aspect of R-symmetric models is the Higgs sector particle
content.  Models have been proposed in the literature with four Higgs
doublets~\cite{Kribs:2007ac}, two Higgs doublets in which the role of
the down type Higgs is played by one of the lepton
doublets~\cite{Frugiuele:2011mh}, one up type Higgs
doublet~\cite{Davies:2011mp} or even with no Higgs doublets at all,
with the role of the Higgs being played by one of the slepton
doublets~\cite{Riva:2012hz}.

As already pointed out, naturalness is among the reasons motivating
the study of models with Dirac gauginos.  However, a solid and
complete statement about the fine-tuning cannot be done without a full
analysis of how a 125 GeV Higgs mass is obtained within this
framework.  The situation has been studied
in~\cite{Benakli:2012cy}, where however the R-symmetric case was not
considered.\footnote{In Ref.~\cite{Chakraborty:2013gea} the question 
of the Higgs mass and fine-tuning is investigated in a scenario with additional right handed neutrinos.}  
This case is going to be the focus of this paper.  As we
will explain, respecting the R-symmetry in the Higgs sector changes
dramatically how the lightest Higgs mass is raised up to 125 GeV (see \cite{Fok:2012fb}).
Indeed, while in~\cite{Benakli:2012cy} this is achieved through an
NMSSM-like tree level enhancement of the Higgs mass, here this
possibility is forbidden by the R-symmetry.\footnote{Another possible 
extension is through new $U(1)$ D-terms, as explored in \cite{Itoyama:2013vxa}. 
This proposal, however, also involves R-symmetry breaking effects.}  However, it turns out
that the extra matter necessary to respect the R-symmetry, \emph{i.e.}
the adjoint scalars and the inert doublets, can provide radiative
corrections comparable to the stop one, giving a 125 GeV Higgs with a
few percent level fine-tuning.

\section{Electroweak symmetry breaking in R-symmetric models}

As already explained, preserving the R-symmetry typically requires an
enlarged Higgs sector.  For definiteness, we will present the
Lagrangian for the four Higgs doublet model~\cite{Kribs:2007ac}, in
which the two doublets with R-charge 0, $H_u$ and $H_d $, acquire a
vev while the two with R-charge 2, $R_u$ and $R_d$, are inert
doublets.  Another, more economical, possibility is to have the
sneutrino as the down type Higgs so that just two doublets, $H_u$ and
$R_d$ are needed.\footnote{It is also possible to have an even more
economical Higgs sector~\cite{Riva:2012hz} where the sneutrino gives
mass to the up type fermions via SUSY breaking Yukawa couplings.
However, in this case the Higgs quartic is generated by SUSY breaking
as well.} We will focus
on the large $\tan{\beta}$ limit (where $\tan\beta \gtrsim 10$) in which most of electroweak symmetry
breaking is through $H_u$, with the extra Higgs states
decoupled from the electroweak symmetry breaking sector. In this limit, we expect the
various models to give similar results.

The superpotential of the model is given by:
\bea
\label{superpotential}  
W &=& W_{Yukawa}+W_{Higgs}  \nn \\ [0.4em]
  W_{Yukawa}&=& H_u Q Y_u u^c
  +  H_d Q Y_d d^c+     H_d L Y_e e^c,  \\ [0.4em]
 \label{eq:W}
 W_{higgs}& = & \sqrt{2} \lambda_T ^u H_u T R_d +\sqrt{2}  \lambda_T ^d  R_u T H_d+\lambda_S^u H_u S R_d +\lambda_S^d R_u S H_d
\nonumber  \\ && \mbox{} +
 \mu_u H_u R_d+\mu_d R_u H_d \, .  \nn 
 \eea
We write the triplet superfield normalized as 
\begin{equation}
 T = \frac{1}{\sqrt{2}}
 \begin{pmatrix}
  T^0 & \sqrt{2} T^+ \\
  \sqrt{2} T^- & - T^0 \\
 \end{pmatrix}\, , 
 \end{equation}
so that the kinetic terms for the (complex) triplet components are
automatically canonically normalized; the factor $ \sqrt{2} $ in front
of $ \lambda_T^i$ is chosen such that $W \supset \lambda_T H_u^0 T^0
R_d^0$.  \\

The R-symmetry allows the gaugino fields $\lambda_i$ to pair up with
the fermionic components of the adjoint superfields, $\psi_i$, through
soft SUSY breaking Dirac masses
\begin{align}
\mathcal{L}_{D}&=  
M_{\tilde B} \lambda_{\tilde B} \psi_{\tilde B}+M_{\tilde W}  \lambda_{\tilde W}^a \psi_{\tilde W}^a+M_{\tilde g} \lambda_{\tilde g} \psi_{\tilde g} + h.c.
\end{align}
Moreover, the soft SUSY breaking scalar terms read
\bea\label{Vsoft}
V_{\rm soft}^{EW} &=& \tilde Q^{\dagger} m^2_{ \tilde Q} \tilde Q+  \tilde u^{\dagger} m^2_{ \tilde u} \tilde u+ \tilde d^{\dagger} m^2_{ \tilde d} \tilde d
+ \tilde L^{\dagger} m^2_{\tilde L} \tilde  L + \tilde e^{\dagger} m^2_{\tilde e} \tilde e +B_{\mu} H_u H_d  
\nonumber \\ 
&& \mbox{} + 
m_{H_u}^2 |H_u|^2  +m_{H_d}^2 |H_d|^2 +m_{R_u}^2 |R_u|^2 + m_{R_d}^2 |R_d|^2 +  m_{s}^2 |S|^2 + m_{T}^2 T^{a \dagger} T^{a}  \nonumber \\ 
&& \mbox{} + 
t_S\, S +  \mbox{}  B_{S} S^2 + \frac{1}{3}A_S\,S^3 + B_{T} T^a T^{a} 
\nonumber \\ 
&& \mbox{} + 
A_{ST}\,ST^2 + A_{SH}\,S H_u H_d + A_{TH}\, H_u T H_d + h.c. 
\eea
We notice that the R-symmetry forbids all the $A$-terms except for
those written above, which together with $t_S$ we will assume to be
negligible for simplicity.\footnote{We can in any case invoke 
a $\mathbb{Z}_2$ parity under which $S$, $T$ and $R_{u,d}$ are odd to forbid these terms.}

Let us now comment on the soft breaking terms in the adjoint sector.
As already explained in the Introduction, Dirac gaugino masses are
generated by supersoft operators and give finite contributions to the
scalar masses.  This property has the beneficial effect of relaxing
the gluino naturalness bound, reducing the tension with the direct
searches~\cite{Kribs:2012gx}.  However not all possible R-invariant
terms that can be constructed out of the adjoint superfields are
supersoft: indeed the non-holomorphic adjoints masses for the singlet
$m_{S}^2$, the triplet $m_T^2$, and the octect $m_O^2$ contribute at
the two-loop level to the $\beta$ functions for the scalar masses,
pushing down their values at low energy.  In particular, a too large
octect scalar mass can eventually induce tachyonic squark masses,
causing charge and color breaking at the weak
scale~\cite{Arvanitaki:2013yja,Csaki:2013fla}.  Furthermore, it is
also important for these three terms (Dirac gaugino masses,
holomorphic and non-holomorphic adjoints scalar masses) to be of the
same order, to avoid tachyons already at tree level.  It turns out,
however, that realizing this spectrum in a UV complete model is quite
challenging.  This resembles the $\mu - B_{\mu }$ problem in gauge
mediation, and it leads to a source of fine-tuning estimated
in~\cite{Csaki:2013fla} to be of order of $0.1 \% $.  In what follows
we will discuss a generic case where also non-holomorphic masses are
present, assuming that the mass hierarchy among the adjoint soft terms
is such as ensure color and charge conservation at the weak scale. 
For definitiveness, we will take the gluino and its scalar octect partner 
to have masses around $4-5$ TeV, \emph{i.e.} large enough to be safe 
from any direct search bound. Moreover, we will assume their ratio to be 
such that the induced tuning on the stop masses is not larger than $20\%$~\cite{Kribs:2012gx}. 
More in general, our attitude towards the bounds on scalars and higgsinos masses 
is such that the LHC constraints can be very mild, 
since our Higgs sector can for example be embedded in a
baryonic RPV scenario~\cite{Frugiuele:2012pe}.

\vspace{2mm} 
The total scalar potential is
\bea
\label{potential}
&&V^{EW} = V_F^{EW} + V_D^{EW} + V^{EW}_{\rm soft} ~,\nn \\
V_F^{EW} &=& \sum_i \left| \frac{\partial W}{\partial \phi_i} \right|^2~,
\hspace{1cm}
V_D^{EW} = \frac{1}{2}\,\sum_{a=1}^3 (D_2^a)^2 + \frac{1}{2}\, D_Y^2~,
\eea
with $W$ defined in Eq.~(\ref{superpotential}) and $V^{EW}_{\rm soft}
$ given in Eq.~(\ref{Vsoft}).  The presence of additional chiral
superfields charged under $SU(2)_L \times U(1)_Y$ modifies the
expression for the D-terms:
\bea \label{eq:D-term}
 D_2^a &=& g \left(H_u^{\dagger}  \tau^a H_u + H_d^{\dagger}  \tau^a H_d + R_u^{\dagger}  \tau^a R_u+ R_d^{\dagger}  \tau^a R_d+ \vec T^\dag \lambda^a \vec T\right) 
 +\sqrt{2} M_{\tilde W} \left(\vec T^a + \vec T^{\dag a} \right)  , \nn \\
 D_Y&=& \frac{g'}{2}\left( H_u^{\dagger} H_u + R_u^{\dagger}  R_u - H_d^{\dagger}  H_d - R_d^{\dagger} R_d \right) 
 + \sqrt{2} M_{\tilde B} \left(S+S^{\dagger}\right), 
 \eea
where $M_{\tilde B}$ and $M_{\tilde W}$ are the Dirac Bino and Wino
masses, $ \tau^a$ and $ \lambda^a$ are the two and three dimensional
$SU(2)$ generators respectively, while $\vec T^a = \sqrt{2} \, {\rm
tr}( \tau^a T )= \left\{ \frac{T^+ + T^-}{\sqrt{2}}, \frac{T^- -
T^+}{\sqrt{2} i}, T^0 \right\}.$\\

Writing the neutral fields as
\be
H^0_{u,d} = \frac{h_{u,d} + i a_{u,d}}{\sqrt{2}}, ~~R^0_{u,d} = \frac{r_{u,d} + i a_{r_{u,d}}}{\sqrt{2}}, ~~ T^0 = \frac{t+i a_t}{\sqrt{2}}, ~~S = \frac{s+i a_s}{\sqrt{2}}\, ,
\ee
the scalar potential for the CP even components reads:
\bea\label{eq:neutr_potential}
V &=& \frac{1}{2} \Big[ \left( m_{H_u}^2 +\mu^2 \right) h_u^2 + \left( m_{R_u}^2 +\mu^2 \right) r_u^2 + \left( m_{H_d}^2 +\mu^2 \right) h_d^2 + 
\left( m_{R_d}^2 +\mu^2 \right) r_d^2  \nn \\
&& \mbox{} - 2 B_\mu h_u h_d + \left( 4 M_{\tilde B}^2 + m_S^2 + 2 B_S \right) s^2  + \left( 4 M_{\tilde W}^2 + m_T^2 + 2 B_T \right) t^2 \Big]  \nn \\
&& \mbox{} + \frac{1}{2} \left[  \sqrt{2} \mu \left( \lambda_S s + \lambda_T t \right) \left( h_u^2 + h_d^2 + r_u^2 + r_d^2 \right)  \right. \nn \\
&& \left. \mbox{} +  \left( g M_{\tilde W} t - g' M_{\tilde B} s\right) \left( h_d^2 + r_d^2 - h_u^2 - r_u^2 \right) \right] 
+ \frac{1}{32} \left( g^2+g'^2\right) \left[ \left(h_u^2 - h_d^2 \right)^2 + \left(r_u^2 - r_d^2 \right)^2 \right]\nn \\
&& \mbox{} + \frac{g^2+g'^2}{16} \left( h_u^2 r_u^2 + h_d^2 r_d^2 \right) + \left(\frac{\lambda_S^2 + \lambda_T^2}{4} - \frac{g^2 + g'^2}{16} \right) \left( h_u^2 r_d^2 + h_d^2 r_u^2 \right)  \nn \\
&& \mbox{} + \frac{\lambda_S \lambda_T}{2} s t \left( h_u^2 + h_d^2 + r_u^2 + r_d^2 \right)  + 
\frac{\lambda_T^2 t^2 + \lambda_S^2 s^2}{4} \left( h_u^2 + h_d^2 + r_u^2 + r_d^2 \right) ,
\eea
where we have assumed for simplicity $
\lambda_S^u=\lambda^d_S=\lambda_S,$ $
\lambda_T^u=\lambda^d_T=\lambda_T,$ $ \mu_u=\mu_d=\mu$ and set
$A_{ST}=A_{SH}=A_{TH}=0$.  The minimization conditions for this
potential are written in the Appendix.  The triplet acquires a vev
which is constrained by EWPM to be $\left| v_T \right| \lesssim 3$
GeV. We will discuss more precisely the bounds from EWPM in
Sec.~\ref{sec:EWPM}.

Inspecting the various contributions, we notice that the D-terms
produce the usual MSSM quartic.  However, the Dirac gaugino masses
contribute to reduce the tree level Higgs mass with respect to the
MSSM. Indeed, $V_D$ contains trilinear interactions between the active
Higgs fields (those participating in EWSB) and the scalar adjoints,
\be
V_D \supset \frac{1}{2}  \left(-g M_{\tilde W}  t + g' M_{\tilde B}  s \right) h_u^2 -\frac{1}{2}  \left( -g M_{\tilde W}  t + g' M_{\tilde B}  s \right) h_d^2 \, , 
\ee
which after EWSB push down the lightest eigenvalue due to mixing.

In addition, the R-symmetry forces the active Higgs fields to couple
only with the inert doublets (those that do not get vevs) and not
among themselves, so that any NMSSM-like quartic term $
\lambda_{S,T}^2 h_u^2 h_d^2$ is forbidden.  As a consequence, the MSSM
tree level upper bound $(m_h)^2_{tree} \leq m_Z^2 \cos^2{2 \beta}$
applies, and the lightest scalar mass is maximized in the large
$\tan\beta$ regime.  The situation is different when the R-symmetry is
broken in the Higgs sector.  In this case $W \supset \lambda_T H_u T
H_d+ \lambda_S H_u S H_d$ and in the low $\tan\beta$ regime the usual
NMSSM-like tree level enhancement is recovered \cite{Benakli:2012cy}.

A more complete discussion of the tree level scalar masses will be
presented in Sec.~\ref{subsec:tree}, where in order to maximize the
lightest eigenvalue we will focus on the large $\tan{\beta}$
regime.\footnote{However, we have checked that it is easy to deform
our benchmark points to obtain examples with moderate $\tan\beta~(\sim 10)$ 
without affecting our conclusions.} In Sec.~\ref{subsec:loop}
we will instead study the loop corrections to the Higgs boson mass.

\begin{figure}[tp]
  \begin{center}
    \includegraphics[height=200 pt]{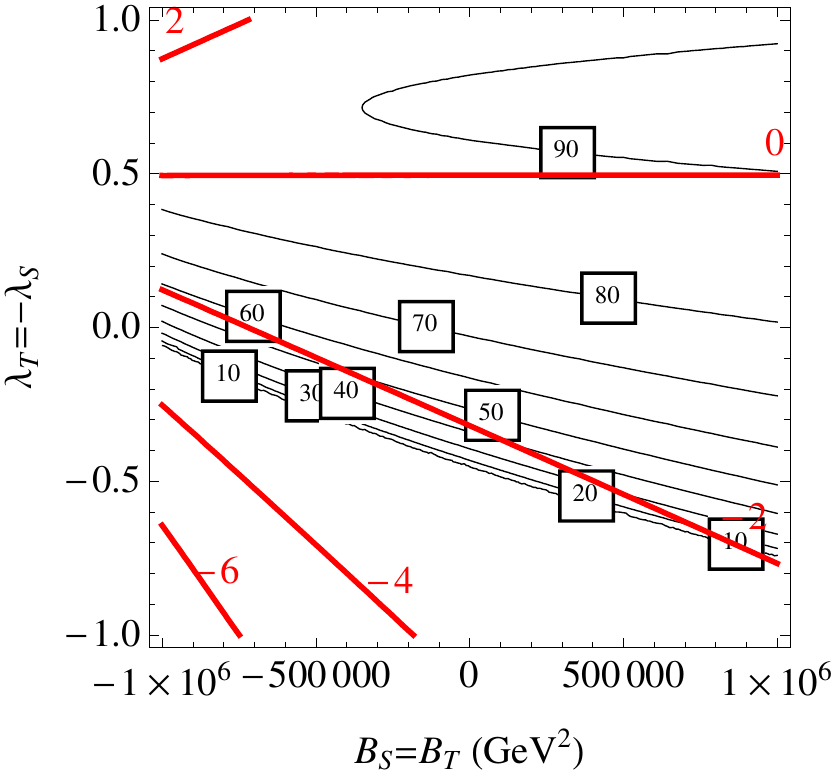} 
    \includegraphics[height=200 pt]{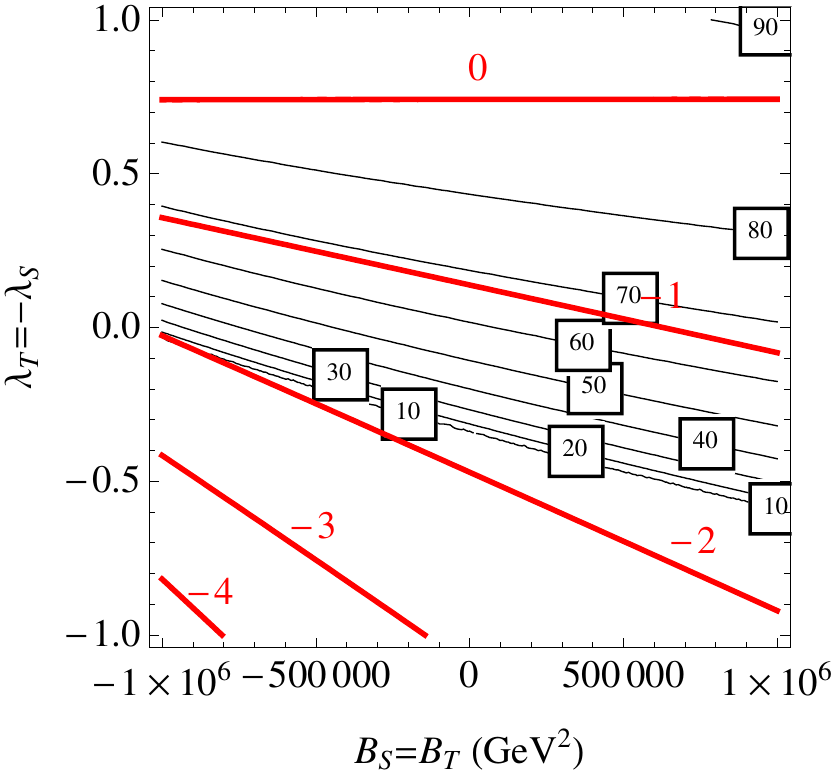}
  \end{center}
\caption{\label{fig:mh_vs} Tree level Higgs boson mass in GeV (black
lines) and singlet vev $v_S$ (red lines) as a function of $B_T = B_S$
and $\lambda_T = - \lambda_S$.  Left: $M_{\tilde W} = M_{\tilde B} =
600$ GeV, $m_T=m_S=1500$ GeV and $\mu = 300$ GeV. Right: $M_{\tilde W}
= M_{\tilde B} = 900$ GeV, $m_T=m_S=1500$ GeV and $\mu = 300$ GeV.}
\end{figure}
%

\subsection{Tree level Higgs mass}\label{subsec:tree}

We have already pointed out that Dirac gaugino masses constitute an
irreducible source of mixing between active Higgs fields and adjoint
scalars.  This push-down effect may in part be kept under control by
the supersymmetric couplings $\lambda_T$, $\lambda_S$ and by the $\mu$
term.  This is evident looking at the off diagonal elements of the
mass matrix for CP even scalars (see Appendix~\ref{app:potential_mass}
for the complete expressions):
\bea\label{eq:mixing}
m^2_{h_u,t} &=& v(-\sqrt{2}  g M_{\tilde W} +2  \lambda_T ( \lambda_S v_S+  \lambda_T v_T+ \mu) )\, , \nn \\
m^2_{h_u,s} &=& v(+\sqrt{2}  g' M_{\tilde B} +2  \lambda_S ( \lambda_S v_S+  \lambda_T v_T+ \mu) )\, .
\eea
Anticipating that $\lambda$ couplings of order one are helpful to
increase the Higgs boson mass at loop level when the stops are not too
heavy, Sec.~\ref{subsec:loop}, and insisting on relatively small $\mu$ values as suggested by
naturalness (see Sec.~\ref{sec:FT}), we see that the terms in
Eq.~(\ref{eq:mixing}) can be kept under control for small singlet and
triplet vevs.  This arises from a partial cancellation between the
first and last terms.  Since by field redefinitions we can always
choose $g > 0$ and $M_{\tilde B}, M_{\tilde W}, \mu >0$, we conclude 
that $\lambda_T>0$ and $\lambda_S<0$ are preferred to obtain
smaller $m^2_{h_u,t}$ and $m^2_{h_u,s}$.  This is confirmed in
Fig.~\ref{fig:mh_vs}, where we show the tree level Higgs boson mass,
together with the singlet vev $v_S$, as a function of $B_T=B_S$ and
$\lambda_T = -\lambda_S$.  In both cases, it is possible to get $m_h
\simeq m_Z$ for small and positive $v_S$.

Going back to the mixing between $h_u$, $s$ and $t$ and the related mass reduction, a simple formula can be obtained 
in the limit of small $v_T$, $v_S$ and large hierarchy between Dirac gaugino masses and non-holomorphic adjoint masses, $M_D \ll m_{adj}$.
For $\tan\beta \gg 1$, the lightest tree level mass is:
\be\label{eq:mh_approx}
(m_h^2)_{tree} \simeq m_Z^2 -v^2 \frac{(-\sqrt{2} g M_{\tilde W} +2 \lambda_T \mu)^2}{m^2_{T_R}}-
v^2 \frac{(\sqrt{2} g' M_{\tilde B} +2 \lambda_S \mu)^2}{m^2_{S_R}}\; ,
\ee
where $m^2_{T_R} = 4 M_{\tilde W} + m^2_T + 2 B_T$ and $m^2_{S_R} = 4 M_{\tilde B} + m^2_S + 2 B_S$ are the masses 
of the real parts of the adjoint scalars before EWSB. Let us stress that the presence of supersymmetric couplings, as well as holomorphic and 
non-holomorphic masses for the adjoint scalars, improves the situation with respect to \cite{Fox:2002bu}, where the quartic coupling vanishes 
for decoupled adjoint scalars (see also \cite{Benakli:2012cy}).

\subsection{Radiative corrections to the Higgs mass}\label{subsec:loop}

We compute now the 1-loop corrected Higgs mass using the
Coleman-Weinberg potential:
\be
V_{Higgs}^{CW}=\frac{1}{64 \pi^2} \left[ \sum_i  (-1)^{2 J_i+1 }\left(2 J_i+1 \right) m_i^4 \left( \log\frac{m_i^2}{Q^2} -\frac{3}{2} \right) \right]\; .
\ee
The sum is to be taken over all the states coupled to the Higgs, with
$m_i^2$'s the field dependent masses.  To obtain analytic expressions
for the loop corrections, we expand the field dependent masses in
powers of $h_u$, setting to zero the singlet and triplet backgrounds
(we know from the previous section that $v_S$ must be small in order
for the tree level Higgs mass not to be too different from $m_Z$,
while $v_T$ must be small to fulfill the precision measurement
constraints).  We do not present here the full analytical expressions,
since they are lengthy and not particularly transparent.  Simple
expressions can be obtained for $ M_D \ll m_{adj}$ or $ m_{adj} \ll
M_D$, where $M_D$ and $m_{adj}$ are common mass scales for Dirac
gauginos and adjoint scalars, respectively.
\begin{description}
\item[Region 1.]  When the scalar CP even and CP odd masses are
significantly larger than the gaugino masses, $\mu \ll M_D \ll m_{adj}
$, we have the following contribution to the Higgs quartic coupling:
\bea\label{eq:CW_quartic}
V_{Higgs}^{CW} &\supset& \frac{1}{4} \left[ \frac{5 \lambda_T^4 + 2\lambda_T^2 \lambda_S^2 + \lambda_S^4}{32 \pi^2} \log\frac{m^2_{R_d}}{Q^2} + 
\frac{\lambda_T^2}{32\pi^2} \left(5 \lambda_T^2 +2 \lambda_S^2 \frac{m_T^2}{m_T^2 - m_S^2}\right) \log\frac{m_T^2}{Q^2} \right.\nn \\  
 && \mbox{} \left.+ \frac{\lambda_S^2}{32\pi^2} \left(\lambda_S^2 - 2 \lambda_T^2 \frac{m_S^2}{m_T^2-m_S^2} \right)\log\frac{m_S^2}{Q^2} 
 - \frac{\lambda_T^2 \lambda_S^2}{16 \pi^2} \right] h_u^4 \nn \\ \nn \\
&& \mbox{} - \frac{1}{4} \left[ \frac{\lambda_T^2}{16\pi^2} \left( 5 \lambda_T^2 + 2 \lambda_S^2 \frac{M^2_{\tilde W}}{M^2_{\tilde W} - M^2_{\tilde B}} \right) \log\frac{M^2_{\tilde W}}{Q^2}
- \frac{\lambda_S^2 \lambda_T^2}{8 \pi^2}  \right.\nn \\
&& \mbox{} \left. + \frac{\lambda_S^2}{16 \pi^2} \left( \lambda_S^2 - 2 \lambda_T^2 \frac{M^2_{\tilde B}}{M^2_{\tilde W} - M^2_{\tilde B}} \right) \log\frac{M^2_{\tilde B}}{Q^2} 
\right] h_u^4 \, ,
\eea
where $ Q$ is the renormalization scale and the first two lines show
the scalar contribution while the third one shows the fermionic one.
We checked that this expression is still a good approximation in the
more interesting limit where milder hierarchies among the masses hold.
However, in what follows we will use the exact expressions to compute
the Higgs boson mass.  A particularly simple expression can be
obtained in the limit $m_{R_d}^2 \simeq m_T^2 \simeq m_S^2 = m^2_{adj}$ and $M_{\tilde W} \simeq M_{\tilde B} = M_D$.  The
Higgs quartic is then
\be\label{eq:CW_quartic2}
V_{Higgs}^{CW} \supset \frac{1}{4} \left[ \frac{5 \lambda_T^4 + 2 \lambda_T^2 \lambda_S^2 + \lambda_S^4}{16\pi^2} \log\frac{m_{adj}^2}{M_D^2} 
+ \frac{\lambda_S^2 \lambda_T^2 }{16\pi^2}\right] h_u^4 \, ,
\ee
so that a relevant positive contribution to the quartic can be
obtained for a large enough ratio $m_{adj}/M_D$.

\item[Region 2.]  In the opposite limit, $ m_{adj} \ll M_D $, the
one-loop contribution to the Higgs quartic is :
\bea
V^{CW}_{Higgs} &\supset & \frac{1}{4} \left[ \frac{5 \lambda_T^4}{32 \pi^2} \log\frac{M_{\tilde W}^2}{Q^2} + 
\frac{\lambda_S^4+ 2 \lambda_T^2 \lambda_S^2}{32 \pi^2} \log\frac{M_{\tilde B}^2}{Q^2} \right] h_u^4  \nn \\
& & \mbox{} - \frac{1}{4} \left[ \frac{5 \lambda_T^4}{16 \pi^2}  \log\frac{M_{\tilde W}^2}{Q^2} +
\frac{\lambda_S^4 }{16\pi^2} \log\frac{M_{\tilde B}^2}{Q^2} -\frac{\lambda_T^2 \lambda_S^2}{8\pi^2} \right] h_u^4
\eea
where we have also assumed $m_{R_d} \ll M_D$.  The first line shows
the scalar contribution, the second line the fermionic one.

Putting all together, we end up with
\be \label{eq:mh_region2}
V_{Higgs}^{CW} \supset \frac{1}{4} \left[ -\frac{5 \lambda_T^4}{32 \pi^2} \log\frac{M_{\tilde W}^2}{Q^2} 
- \frac{\lambda_S^4-2\lambda_T^2 \lambda_S^2}{32\pi^2} \log\frac{M_{\tilde B}^2}{Q^2} +\frac{\lambda_T^2 \lambda_S^2}{8\pi^2} \right] h_u^4
\ee
so that we expect this contribution to be always negative.  Let us
notice that this region corresponds to the pure supersoft spectrum,
where indeed the non-holomorphic scalar masses are negligible and $
M_D^2 \gtrsim B$ in order to avoid problems with tachyonic masses.
Furthermore, $ m^2_{R_d}$ is given by the gaugino induced one-loop
correction~\cite{Fox:2002bu}:
\be\label{eq:mRd}
m^2_{R_d} = \frac{ \alpha_2}{\pi} M^2_{\tilde W} \log{\frac{ 4 M^2_{\tilde W} - 2 B_T}{ M_{\tilde W^2}}} \, ,
\ee 
with an inert doublet therefore too light to give any significant
boost to the Higgs mass.
\end{description}
For comparison, the well known stop contribution is given
by~\cite{Carena:1995wu}
\bea\label{eq:lambda_stop}
V_{Higgs}^{CW} &\supset& \frac{1}{4} \left[ \frac{3}{16 \pi^2} y_t^2 \left(y_t^2-  \frac{m_Z^2}{2v^2}\right) \log\frac{M^2}{m_t^2}  \right. \nn\\
&& \left. \mbox{} + \frac{3 y_t^4}{(16\pi^2)^2} \left(\frac{3}{2} y_t^2 - 32 \pi \alpha_3(m_t) \right) \log^2\frac{M^2}{m_t^2}   \right] h_u^4 \, ,
\eea
where $y_t$ is the top Yukawa coupling, $\alpha_3$ the strong coupling
constant and $M$ is a common soft SUSY breaking stop mass scale.  We
show also the two-loop contribution, since the term proportional to
the strong gauge coupling may reduce in a significant way the Higgs
quartic.

\vspace{2mm} The simplified expressions, Eqs.~(\ref{eq:CW_quartic}),
(\ref{eq:CW_quartic2}), suggest that for $|\lambda_T| \simeq |\lambda_S|
\simeq y_t$ the new states may give a contribution comparable to the
stop one, depending on the mass hierarchy.  In this sense, they can be
regarded as ``additional stops'', which in principle can allow for a
collective loop enhancement of the Higgs quartic.  Moreover, since
both the triplet and the singlet are uncolored, we do not expect the
two loop terms proportional to the gauge couplings to give a reduction
analogous to the one proportional to $\alpha_3$ in the stop sector,
making more effective the loop boost achieved through these states. 
Whether or not this scenario will allow to obtain a $125$ GeV Higgs
with less fine-tuning than in the MSSM will be studied in detail in
Sec.~\ref{sec:FT}.

\vspace{2mm}  As last comment, let us notice that the condition 
$|\lambda_T| \simeq |\lambda_S| \simeq y_t$ (\emph{i.e.} rather large 
values for the trilinear couplings at the weak scale) may imply 
a loss of perturbativity at relatively low scales. Solving 
the RGE's~\cite{Goodsell:2012fm} requiring $\lambda_T=1=-\lambda_S$ 
at the weak scale, we find that the coupling that 
runs faster, $\lambda_T$, reaches $\sqrt{4\pi}$ for scales above $100$ TeV.

\section{Electroweak Precision Measurements}\label{sec:EWPM}

%
\begin{figure}[t]
 \begin{center}
  \includegraphics[scale=0.65]{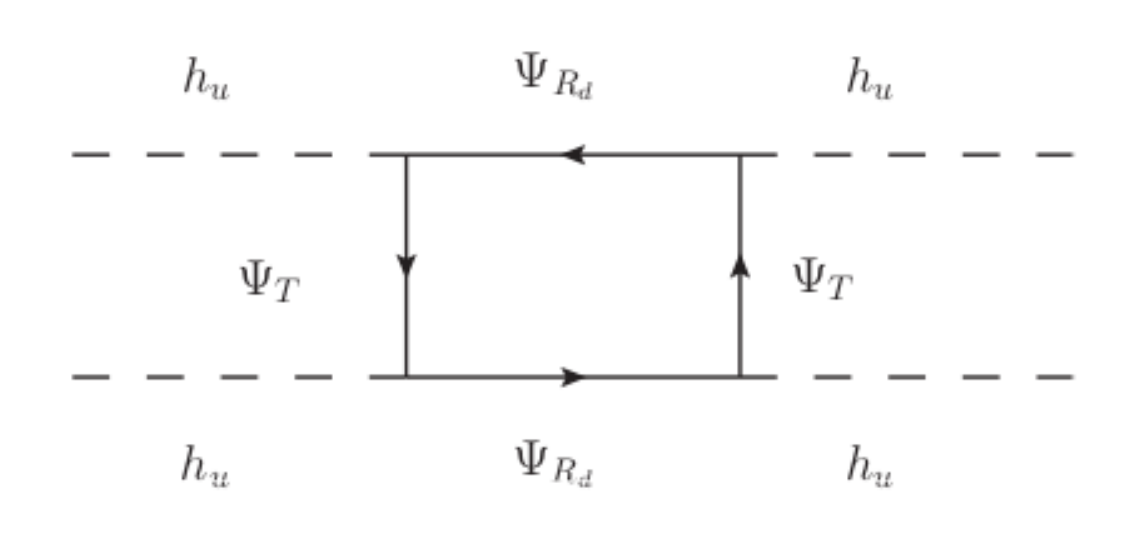}
 \end{center}
 \caption{ \label{fig:box} Box diagram that contributes to the $T$ parameter.}
\end{figure}
Getting a significant help from the triplet and the singlet to raise
the Higgs mass through radiative corrections requires an appreciable
value for the couplings $\lambda_T$ and/or $\lambda_S$.  However, the
same couplings contribute to the $T$ parameter at loop level.
Therefore there exist a potential tension between generating a large
Higgs mass and electroweak precision data.  Besides the loop-level
corrections to $T$ there is already at tree level a dangerous effect
due to the vev for the triplet $T^0$, which can lead to a large
contribution to the $\hat{T}$ parameter (with the standard
$T=\hat{T}/\alpha$):
\be\label{eq:vT}
\hat{T} =  4\frac{v_T^2}{v^2} ,
\ee
which constrain the triplet vev  to be $|v_T| \lesssim 3$ GeV, where
\be
v_T=\frac{\sqrt{2} g M_{\tilde W} -2 \lambda_S \lambda_T v_S -2 \lambda_T \mu }{4 B_T+8 M_{\tilde W}^2+2 m_T^2+ 2 \lambda_T^2 v^2} v^2 \; \; .
\ee
It can be minimized by taking $m_T$ large, or otherwise arranging for
the numerator to be small.  Besides the tree level contributions,
there are contributions coming from loop of superpartners.  A detailed
study of all these contributions will be presented in \cite{Beauchesne:2014pra},
but the dominant effect comes from contributions to $\hat{T}$ from
loops involving the fermionic part of the superfield $H_u$, $T$, $R_d$
and $S$.  Integrating them out at loop level, trough the diagram of
Fig.~\ref{fig:box} lead to the higher-dimension operator associated
with $\hat{T}$:
\be
\frac {\left| H_u^{\dagger} D_\mu H_u\right|^2}{\Lambda^2}~,
\ee
\begin{figure}[t]
  \begin{center}
   \includegraphics[width=.4\textwidth]{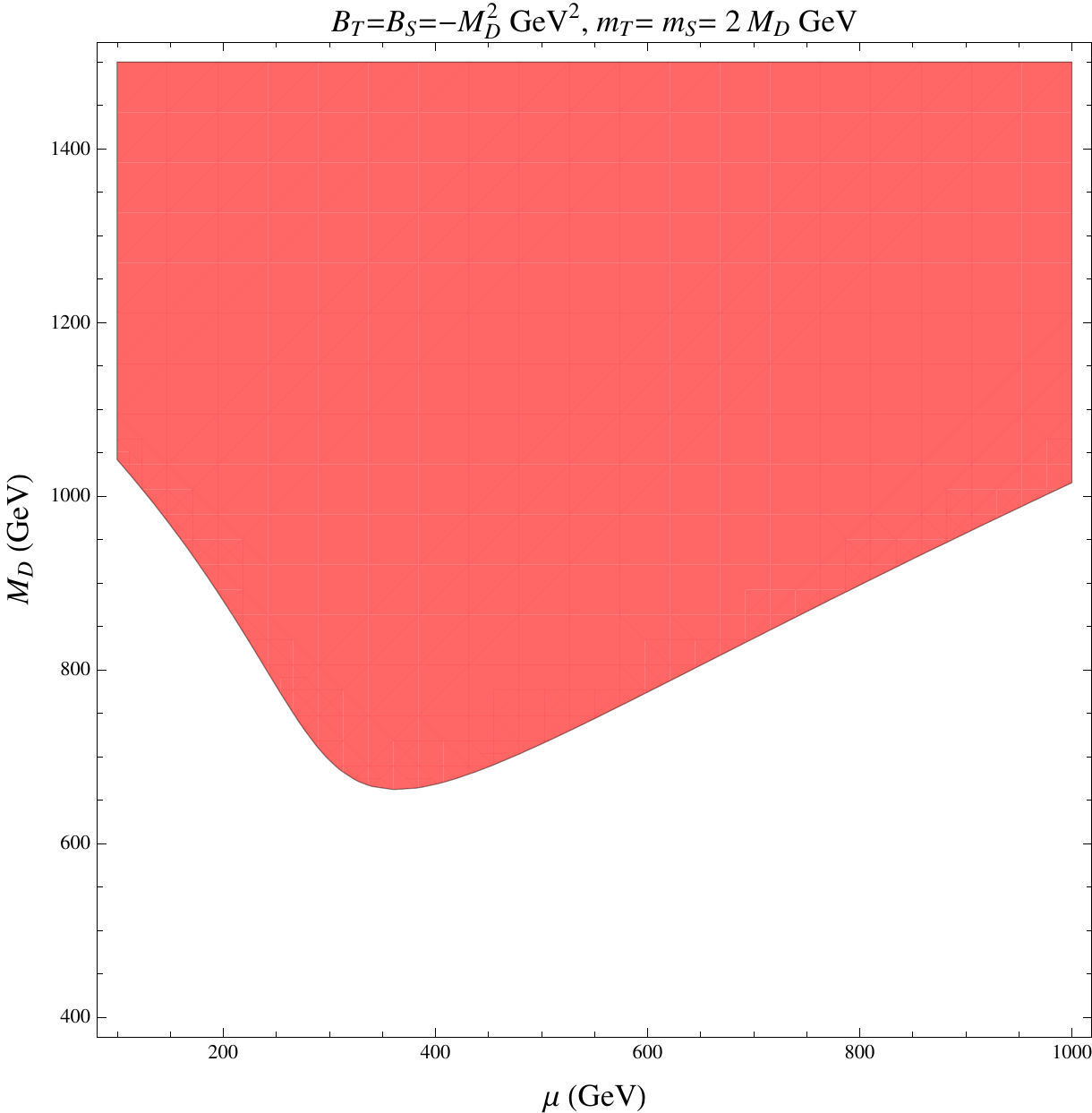}
  \end{center}
  \caption{ \label{fig:EWPM_mu_MD} 
  Region allowed at $95\%$ C.L. by EWPM ($T < 0.2$) as a function of 
  $M_D=M_{\tilde W} = M_{\tilde B}$ and $\mu$. The adjoint holomorphic and non-holomorphic 
  masses are fixed to 
  $m_S = m_T = 2 M_D$ and $B_S = B_T = -M_D^2$, respectively. The supersymmetric trilinear couplings are 
  $\lambda_T = 1 = -\lambda_S$.}
\end{figure}
with a coefficient proportional to $\lambda_T^4$.  Thus the same
coupling which can help to make the Higgs heavier will also lead to
too large contributions to $T$.  To estimate the region of parameter
space excluded by electroweak precision data we compute the $\hat{T}$
parameter due to $v_T$, and from loops of the scalar and fermonic
sector of $H_u,T,R_d$ and $S$.  Imposing $T<0.2$
\cite{Beringer:1900zz}, we find that the fermion loops will force
$\lambda_T \lesssim 1$, whenever $M_D \lesssim 1$ TeV.
We notice however that, as can be anticipated from Eq.~(\ref{eq:vT}), 
the $\mu$ parameter plays an essential role in keeping 
the contributions to $\hat{T}$ under control, at least at tree level. 
This is confirmed in Fig.~\ref{fig:EWPM_mu_MD}, 
from which it can be clearly seen 
that the expected lower bound on $M_D$ is dramatically modified in the 
$\mu = (300-400)$ GeV region, where gaugino masses as low as 
$M_D \simeq 650$ GeV are allowed. This can be essentially traced back to the 
smallness of the numerator of Eq.~(\ref{eq:vT}), that makes the tree level 
contribution to $\hat{T}$ basically negligible.

\section{125 GeV Higgs boson and fine-tuning}\label{sec:FT}

We are now in the position to analyze the region in parameter space in
which not only a $125$ GeV Higgs boson mass can be obtained, but also
the contributions to the $T$ parameter can be kept under control.
Before doing so we comment on the fine-tuning in our set up.  

Following~\cite{Gherghetta:2012gb}, we consider two possible sources of tuning. The first one measures the sensitivity 
of the vev on the fundamental parameters $a_i= \left\{m_T^2 \right.$, $m_S^2$, $m_{R_d}^2$, $m_{Q_3}^2$, $m^2_{u_3}$, $\mu$, 
$M_{\tilde W}$, $M_{\tilde B}$, $B_T$, $B_S$, $\lambda_T$, $\left. \lambda_S \right\}$,
\begin{equation}\label{eq:FT_delta}
\Delta = {\rm max}_{a_i} \left|\frac{\partial \log m_Z^2}{\partial \log a_i}\right|~.  
\end{equation}
The second one measures the sensitivity of the physical Higgs mass $m_h^2$ on the same set of parameters (this time for fixed vev's):
\begin{equation}\label{eq:FT_deltamh}
\Delta_h = {\rm max}_{a_i} \left|\frac{\partial \log m_h^2}{\partial \log a_i}\right|~.
\end{equation}
Let us notice that Eq.~(\ref{eq:FT_deltamh}) effectively measures the tuning on the Higgs quartic coupling.

As customary, the dependence of $m_Z^2$, Eq.~(\ref{eq:minimization}), on the high energy parameters (defined at the scale 
$\Lambda$ at which they are generated) is taken into account solving the RGE's for $m^2_{H_u}$~\cite{Goodsell:2012fm}. 
In the leading-log approximation,
\begin{equation}
 \delta m_{H_u}^2 \simeq \frac{1}{8 \pi^2} \bigg[\left(\lambda_S^2 + 3 \lambda_T^2 \right) m^2_{R_d} + \lambda_S^2 m_S^2 + 3 \lambda_T^2 m_T^2 + 
 3 y_t^2 \left( m^2_{Q_3} + m^2_{u_3} \right) \bigg] 
 \log\frac{\Lambda}{{\rm TeV}}
\end{equation}
In contrast to what happens in models with Majorana gaugino masses,
no dependence on the Dirac gaugino masses is present in the RGE's for 
$m_{H_u}^2$.\footnote{There is nonetheless a dependence
through the finite one-loop contribution analogous to
Eq.~(\ref{eq:mRd}); we checked however that the tuning due to this
contribution is never the dominant one in the interesting regions.} 

Regarding the computation of the fine-tuning, the usual estimates are not valid in this case. Indeed, once $v_T$ and $v_S$ are 
inserted in the expression for $m_Z^2$, Eq.~(\ref{eq:minimization}), 
the dependence on the parameters is different from the MSSM one, which is recovered only in the $m_{adj} \rightarrow \infty$ limit. 
Although the expressions are quite involved, we can obtain a good approximation solving perturbatively the minimum equations in powers 
of $\alpha_i = v^2/(2 B_i + 4 M_i^2 + m^2_i+\lambda_i^2 v^2)$, 
which are small quantities in the region of parameter space we are going to consider (\emph{i.e.} $M_D, m_{adj} \gtrsim 500$ GeV). 
For instance, to order $\alpha_{s,t}$, the triplet and singlet vev's read
\begin{equation}\label{eq:vSvT}
  v_T = \alpha_T \left(\frac{g M_{\tilde W}}{\sqrt{2}} - \lambda_T \mu \right), ~~~~~~
  v_S = -\alpha_S \left(\frac{g' M_{\tilde B}}{\sqrt{2}} + \lambda_S \mu \right), 
\end{equation}
from which it is clear that both quantities are small. As already seen, this implies that for $\lambda_T>0$ and $\lambda_S<0$, 
the tree level mixing between $h_u$ and $s$, $t$ is small, Eq.~(\ref{eq:mixing}). 
Computing the variation of the first of Eqs.~(\ref{eq:minimization}) we obtain
\begin{equation}\label{eq:FT_upperbounds}
 \begin{array}{ccl}
  \Delta_\mu &=& \left|\frac{4 \mu^2}{m_Z^2} + \frac{8 \mu}{m_Z^2} \left[ \lambda_S v_S + \lambda_T v_T + \frac{2 \mu}{m_Z^2} \left( \frac{v_S^2}{\alpha_S} + \frac{v_T^2}{\alpha_T} \right) \right] \right|\, ,\\[0.2cm]
  \Delta_{M_{\tilde{W}}} &=& \left| \frac{4\sqrt{2} g M_{\tilde W} v_T}{m_Z^2} \right| \, ,\\[0.2cm]
  \Delta_{M_{\tilde{B}}} &=& \left| \frac{4\sqrt{2} g' M_{\tilde B} v_S}{m_Z^2} \right| \, ,\\[0.2cm]
  \Delta_{B_T} &=& \left| \frac{8 B_T}{m_Z^2} \frac{v_T^2}{v^2} \right|\, , \\[0.2cm]
  \Delta_{B_S} &=& \left| \frac{8 B_S}{m_Z^2} \frac{v_S^2}{v^2} \right|\, , \\[0.2cm]
  \Delta_{m^2_{R_d}} &=& \left| \frac{\lambda_S^2 + 3 \lambda_T^2}{4 \pi^2} \frac{m^2_{R_d} \log\frac{\Lambda}{{\rm TeV}}}{4m_Z^2} \left[ 1+ \frac{4}{m_Z^2} \left( \frac{v_S^2}{\alpha_S} + \frac{v_T^2}{\alpha_T} \right) \right] \right|\, ,\\[0.2cm]
  \Delta_{m^2_T} &=& \left| \frac{3 \lambda_T^2 m_T^2\log\frac{\Lambda}{{\rm TeV}} }{4\pi^2 m_Z^2} \left[1+\frac{4}{m_Z^2} \left(  \frac{v_S^2}{\alpha_S} + \frac{v_T^2}{\alpha_T} \right) \right] \right|\, ,\\[0.2cm]
  \Delta_{m^2_S} &=& \left| \frac{ \lambda_S^2 m_S^2\log\frac{\Lambda}{{\rm TeV}} }{4\pi^2 m_Z^2} \left[1+\frac{4}{m_Z^2} \left(  \frac{v_S^2}{\alpha_S} + \frac{v_T^2}{\alpha_T} \right) \right] \right|\, ,\\[0.2cm]
  \Delta_{m^2_{Q_3}, m^2_{u_3}} &=& \left| \frac{ 3 y_t^2 m_{Q_3, u_3}^2\log\frac{\Lambda}{{\rm TeV}} }{4\pi^2 m_Z^2} \left[1+\frac{4}{m_Z^2} \left(  \frac{v_S^2}{\alpha_S} + \frac{v_T^2}{\alpha_T} \right) \right] \right|\, .\\[0.2cm]
 \end{array}
\end{equation}
We immediately see that in the limit $v_S = 0 = v_T $ we get the same expressions we 
would have obtained from the MSSM's minimum condition, \emph{i.e.} only $\Delta_\mu$, and 
$\Delta_{m^2_{Q_3}, m^2_{u_3}}$ do not vanish. Switching on the singlet and triplet contributions, 
we get corrections which start at ${\cal O}(\alpha_{T,S})$ (notice that, according to Eq.~(\ref{eq:vSvT}), 
$v_S$ and $v_T$ are ${\cal O}(\alpha_{T,S})$, 
and so are $v_S^2/\alpha_S$ and $v_T^2/\alpha_T$). 
In particular, since $\Delta_{M_{{\tilde W}, {\tilde B}}} \sim {\cal O}(\alpha_{T,S})$ and 
$\Delta_{B_T, B_S} \sim {\cal O}(\alpha^2_{T,S})$, the related tuning is never relevant.

Among the tunings due to the soft SUSY breaking masses, for $\lambda_T \sim \lambda_S \sim 1$ we get comparable results from $m^2_{R_d}$, $m^2_T$ and 
$m^2_{Q_3, u_3}$ (with a slightly worse sensitivity due to the inert doublet mass).
Indeed, taking all the soft masses to be of the same order, we get
\begin{equation}\label{eq:tuning_deg_masses}
 \Delta_{m^2_{R_d}} \simeq 4 \Delta_{m^2_{S}} \simeq \frac{4}{3} \Delta_{m^2_{T}} \simeq \frac{4}{3} \Delta_{m^2_{Q_3}}\, .
\end{equation}
In particular, there is no worsening in the fine-tuning for $m_{\tilde t} \simeq m_{adj}$. 

As usual, we need also to keep under control the tree level tuning due to $\mu$, whose contribution may rapidly become the dominant one (especially in the 
region with relatively small $m_{adj}$ in which the remaining sensitivities are not particularly severe). In particular, 
in the region $M_D, m_{adj} \lesssim 2$ TeV, we have 
\begin{equation}
 \begin{array}{ccl}
  \mu = 100\, {\rm GeV} & \rightarrow & \Delta_\mu \sim 5-10 \, ,\\
  \mu = 200\, {\rm GeV} & \rightarrow & \Delta_\mu \sim 20-30 \, ,\\
  \mu = 300\, {\rm GeV} & \rightarrow & \Delta_\mu \sim 45-55 \, .\\
 \end{array}
\end{equation}
Since $\mu$ gives the Higgsino mass, we need to 
be careful with the limits imposed by direct searches and EWPM. In principle, we would expect smaller values of $\mu$ to be preferred since
they give smaller $\Delta_\mu$. However, as shown in Fig.~\ref{fig:EWPM_mu_MD}, light Higgsinos require heavier gauginos to be compatible with EWPM 
(unless we take $\mu=(300-400)$ GeV). 
As we are going to see, this in turn implies heavier scalars to accommodate 
$m_h = 125$ GeV, with a general worsening of the fine-tuning.

Let us now discuss the sensitivity of the Higgs quartic couplings on the parameters, Eq.~(\ref{eq:FT_deltamh}). 
Integrating out the heavy fields in Eq.~(\ref{eq:neutr_potential}) we obtain 
\begin{equation}\label{eq:effective_quartic}
 V \supset \frac{1}{4v^2} \left[ \frac{m_Z^2}{4} - \left( \frac{v_T^2}{\alpha_T} + \frac{v_S^2}{\alpha_S} \right) +\lambda_{loop} \right] h_u^4
\end{equation}
from which we easily compute
\begin{equation}\label{eq:Deltah}
 \begin{array}{lcl}
  \Delta_h^{(\mu)} &=& \frac{8 \mu}{m_h^2} \left( \lambda_T v_T + \lambda_S v_S \right) + \left. \Delta_h^{(\mu)}\right|_{loop} \, , \\[0.2cm]
  \Delta_h^{(M_{\tilde W})} &=& \frac{4 M_{\tilde W} v_T }{m_h^2} \left( \frac{8 M_{\tilde W} v_T}{v^2} -\sqrt{2} g \right) + \left. \Delta_h^{(M_{\tilde W})}\right|_{loop} \, , \\[0.2cm]
  \Delta_h^{(M_{\tilde B})} &=& \frac{4 M_{\tilde B} v_S }{m_h^2} \left( \frac{8 M_{\tilde B} v_S}{v^2} -\sqrt{2} g' \right) + \left. \Delta_h^{(M_{\tilde B})}\right|_{loop} \, , \\[0.2cm]
  \Delta_h^{(m^2_T)} &=& \frac{4 m^2_T v_T^2}{m_h^2 v^2} + \left. \Delta_h^{(m^2_T)}\right|_{loop} \, , \\[0.2cm]
  \Delta_h^{(m^2_S)} &=& \frac{4 m^2_S v_S^2}{m_h^2 v^2} + \left. \Delta_h^{(m^2_S)}\right|_{loop} \, . \\[0.2cm]
 \end{array}
\end{equation}
\begin{figure}[t]
  \begin{center}
  \begin{tabular}{cc}
   \includegraphics[width=.47\textwidth]{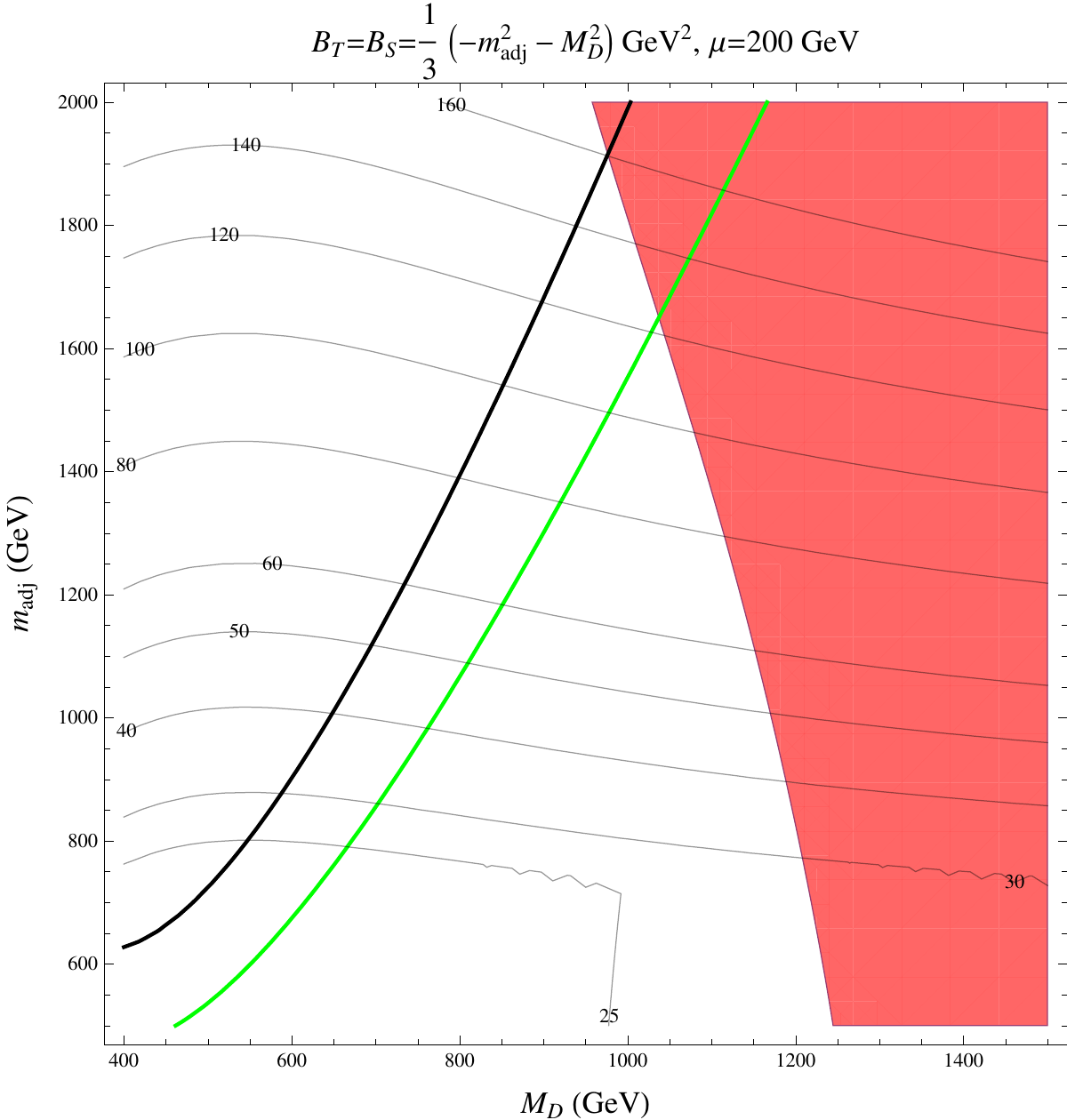} & \includegraphics[width=.47\textwidth]{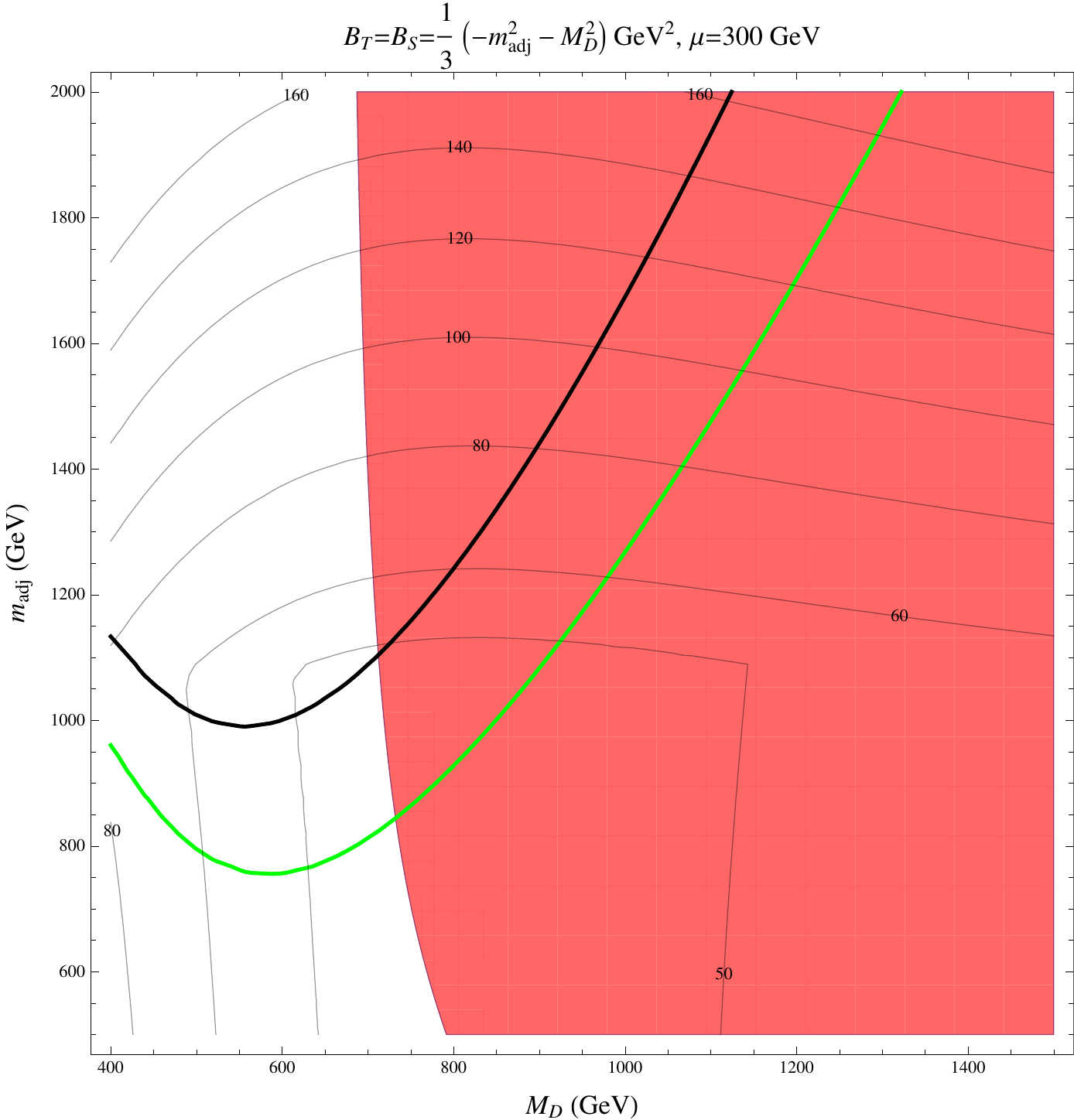}
   \end{tabular}
  \end{center}
  \caption{ \label{fig:fin1} Higgs boson mass $m_h=125$ GeV (black and green thick
  lines) and fine-tuning parameter $\Delta$ (thin lines), as a function of
  $M_D=M_{\tilde W} = M_{\tilde B}$ and $m_{adj} = m_T = m_S=m_{R_d}$, 
  for $B_T=B_S= - \frac{1}{3} (m^2_{Adj}+M_D^2) $.  We fix
  $\lambda_T = 1 = -\lambda_S$. The upper (black) curve refers to a common stop
  mass of $m_{stop} = 300$ GeV, the lower (green) curve to $m_{stop} = m_{adj}$. Left panel: $\mu=200$ GeV; right panel: $\mu=300$ GeV. 
  The red region is allowed at $95\%$ C.L. by EWPM ($T < 0.2.$) }
\end{figure}
The contributions dubbed $\left.\Delta_h^{(i)}\right|_{loop}$ are those coming from $\lambda_{loop}$ in Eq.~(\ref{eq:effective_quartic}).
From Eq.~(\ref{eq:Deltah}) it is clear that all the tree level contributions start either at ${\cal O}(\alpha_{T,S})$ or at ${\cal O}(\alpha^2_{T,S})$ 
and are thus going to be irrelevant. It can nevertheless happen that some or all of the 
$\left.\Delta_h^{(i)}\right|_{loop}$ are large.
We can get an idea of the typical tuning arising at loop level 
using the approximation of Eqs.~(\ref{eq:CW_quartic2}) together with the stop contribution, Eq.~(\ref{eq:lambda_stop}): we always have 
$\left.\Delta \right|_{loop} \lesssim 1$. We checked numerically that this is also the case when the complete loop contributions are taken into account, 
so that in the following we will discard $\Delta_h$.\\

Our main results are shown in Fig.~\ref{fig:fin1}.  
We present the results as a function of
$M_D=M_{\tilde W} = M_{\tilde B}$ and $m_{adj} = m_T = m_S=m_{R_d}$,
with couplings fixed to $\lambda_T = 1 = -\lambda_S$. On the left panel $\mu=200$ GeV, 
while on the right panel $\mu=300$ GeV. 
The solid thick lines correspond
to $m_h = 125$ GeV, with the red region allowed at $95\%$ C.L. by
EWPM. We also show the largest among the fine-tuning parameters $\Delta_i$ (thin black
lines), Eq.~(\ref{eq:FT_upperbounds}), fixing $\Lambda = 20$ TeV.\footnote{In this work
we assume that the required soft parameters can be obtained naturally
with the appropriate values at this scale.  However, as mentioned
previously, this is a somewhat non-trivial task and could be the
source of additional fine-tuning~\cite{Arvanitaki:2013yja}.} 
The $m_h=125$ GeV thick lines refer 
to two different stop masses: $m_{\tilde t} =300$ GeV~\footnote{A detailed
study of the LHC phenomenology of the model is outside the scope of
the present work, therefore we assume $ m_{\tilde t } \sim 300 $ GeV
to be still allowed by the LHC either because of a very compressed spectrum or because 
of baryonic R-Parity violating couplings in the
superpotential.} for the upper
curve and $m_{\tilde t} = m_{adj}$ for the lower curve. 

Let us comment on two counterintuitive features of our results: the correct Higgs mass is achieved with 
less fine-tuning for \emph{heavier} stops and for \emph{heavier} Higgsinos. This can be understood as 
follows: for the upper (black) curves, the lightness of the stops is such that the main boost to 
the Higgs quartic comes from the adjoint and inert fields. On the contrary, for the lower (green) curves 
the stop boost to the Higgs quartic is relevant. However, as already pointed out, there is no 
worsening in the tuning for $m^2_{\tilde stop} = m^2_T = m^2_{R_d}$, Eq.~(\ref{eq:tuning_deg_masses}). Moreover, 
the ``collective'' quartic enhancement in the lower curves allows for smaller soft SUSY breaking masses, implying thus less tuning.
Turning to the $\mu$ parameter, we already observed that compatibility with EWPM for lighter 
Higgsinos require heavier gauginos (\emph{i.e.} larger $M_D$). In addition, it is clear from the shape of the Higgs mass curves 
in Fig.~\ref{fig:fin1} that this in turn requires heavier scalars to get $m_h = 125$ GeV, so that a worsening in the tuning 
is expected. This is indeed the case in Fig.~\ref{fig:fin1}: for $\mu=300$ GeV compatibility between $m_h=125$ GeV and EWPM 
is achieved for $m_{adj} \gtrsim 800 - 1100$ GeV (for $m_{\tilde t} = m_{adj}$ or $300$ GeV, respectively), \emph{i.e.} when the sensitivity 
is still dominated by $\mu$. On the contrary, for $\mu=200$ GeV the scalar masses are pushed up to $m_{adj} \gtrsim 1500-1900$ GeV 
(again for $m_{\tilde t} = m_{adj}$ or $300$ GeV, respectively), in a region in which the soft SUSY breaking masses dominate the tuning.

For comparison, in the MSSM with maximal stop mixing $m_h=125$ GeV is 
achieved with $\Delta \gtrsim 100-200$~\cite{Hall:2011aa}, while since 
for $A_t = 0$ stop masses around $10$ TeV are needed~\cite{Bagnaschi:2014rsa}, 
we can estimate $\Delta \gtrsim 2000$. We do not attempt here to follow~\cite{Gherghetta:2012gb} and find 
the minimum tuning achievable in the model; it is however clear that among the beneficial effects 
of our R-symmetric scenario we have a significant fine-tuning reduction (modulo potential 
additional sources from the UV that might be reduced with further model building).\\
\begin{figure}[t]
  \begin{center}
   \includegraphics[width=.6\textwidth]{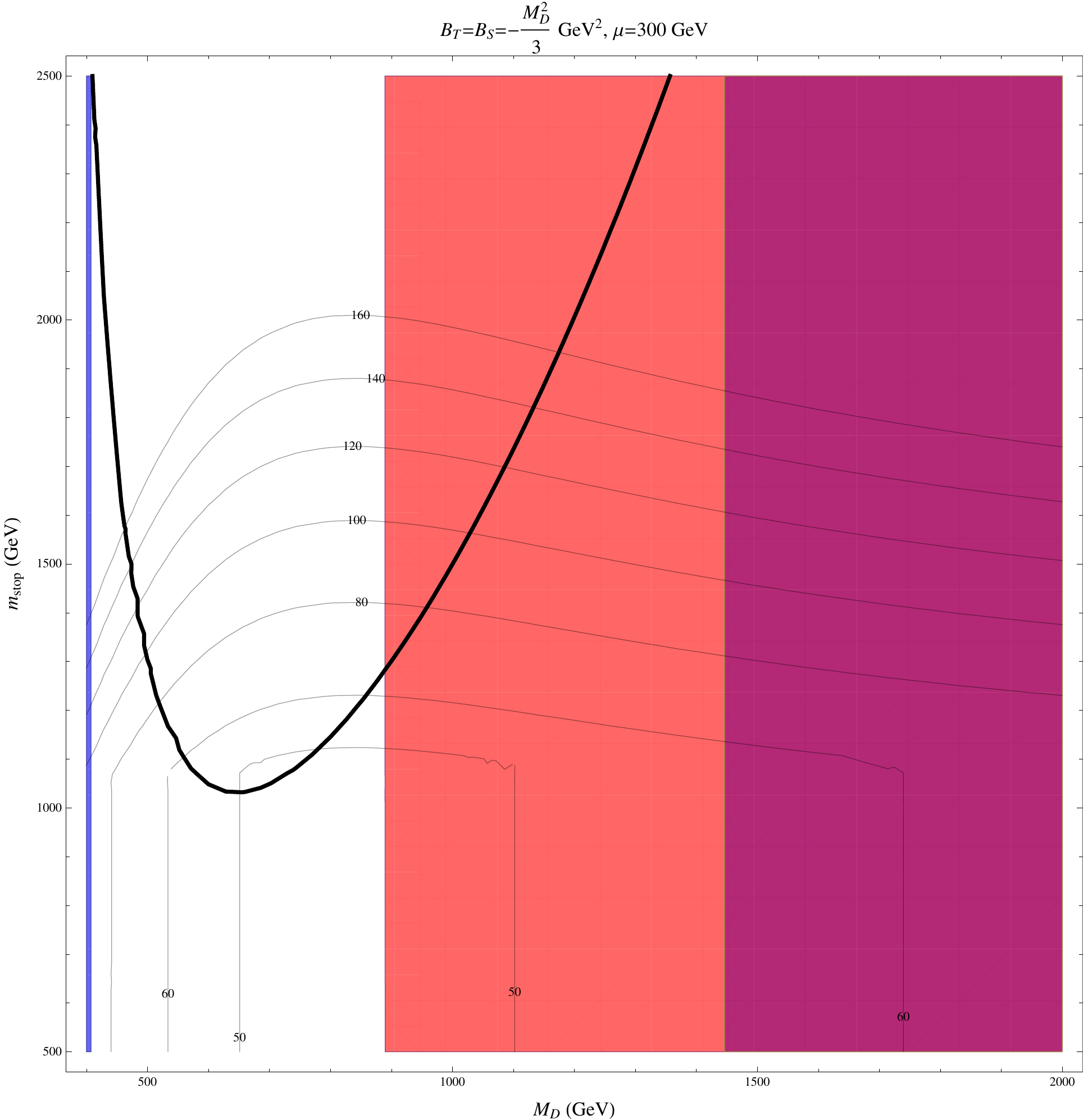}
    \end{center}
  \caption{ \label{fig:SS1} Higgs boson mass $m_h=125$ GeV (solid
  line) and fine-tuning parameter (thin lines), as a function of
  $M_D=M_{\tilde W} = M_{\tilde B}$ and $m_{\tilde t}=m_{R_d}$, with
  $m_T=m_S=0$ (supersoft limit), and $B_T=B_S= - \frac{M_D^2}{3} $.
  We fix $\lambda_T = 1 = -\lambda_S$ and $\mu= 300$ GeV. Blue region:
  spontaneous charge and/or CP breaking.  Red region: allowed at
  $95\%$ C.L. by EWPM ($T < 0.2.$) .  Purple region: $m_{\tilde
  \ell_R}>100$ GeV}
\end{figure}

As a final comment, we consider the case of a pure supersoft spectrum,
\emph{i.e.}~the case in which we set the non-holomorphic adjoint
masses $m_T$ and $m_S$ to zero.  With $ m^2_{R_d}$ given
by~(\ref{eq:mRd}), we easily see from Eq.~(\ref{eq:mh_region2}) that
the only relevant radiative correction comes from the stop sector,
with the gaugino contributions decreasing the Higgs mass.  However, it
is easy to imagine that the full UV completion of the model may contain a
sector which couples the Higgs multiplets to messenger fields.  This
can generate the required $\mu $ and $ B_{\mu}$ terms, as well as
extra contributions to the $R_d$ mass which may then push up the Higgs
mass even with moderately light stops.  In any case, relying only on
the loop corrections from $R_d$ makes the boost to the Higgs quartic
less efficient than in the non-supersoft case, and this makes the
model less natural, since $m^2_{R_d}$ is associated with a larger
fine-tuning.  There is also another important constraint to take into
account in a pure supersoft spectrum: all the sfermions acquire mass
via finite gaugino one loop contribution and are therefore predicted
(modulo the running from the adjoint mass scale down to the weak
scake) in term of $M_D$.  This tends to make the sleptons, especially
the right-handed ones which only have hypercharge gauge couplings,
quite light.  This is illustrated in Fig.~\ref{fig:SS1}, where we show
once again, in the $M_D-m_{\rm stop}$ plane, the line
corresponding to a 125 GeV Higgs together with the region allowed by
EWPM (in red).  We also show the region where the slepton has a mass
greater than $100$ GeV (purple region), which correspond to the LEP
bound.  We see that the slepton constraint pushes all the masses to be
very heavy, into a region with very large fine-tuning.  This leads to
the conclusion that sizable non-holomorphic adjoints masses are
required to reduce the fine-tuning.

\section{Conclusions} 

We are finally in an era in which experiments are directly exploring
the electroweak scale, and will (at least in part) shed light on
whether or not the electroweak scale is natural.  The first LHC run
has already provided us with some indications.  The general message
seems to be that our simplest natural models are by now tuned at the
percent level, or worse.  While it can turn out that this is the level
of tuning of the EW scale, it may also be taken to motivate the search
for more natural (although less minimal) models.  One such possibility
is the supersymmetric model with a quasi exact $U(1)_R$ symmetry
considered in this work.  This kind of models are more natural with
respect to the MSSM, since there is no gluino induced one-loop
contribution to the squark masses.  Moreover, the usual supersymmetric
flavor problem is greatly ameliorated.  The point on which we focus
here is that the adjoint superfields needed to write Dirac gaugino
masses may give relevant loop corrections to the Higgs boson mass:
they act effectively as ``additional stops'', at least in part of the
parameter space.  A possible drawback is that the very same couplings
that help increasing the Higgs boson mass break custodial symmetry,
potentially leading to large contributions to the electroweak
precision measurements.  Our main results are summarized in
Fig.~\ref{fig:fin1}, in which we show the region in
parameter space in which a 125 GeV Higgs mass can be obtained in a way
compatible with EWPM. We also presented on the same plot the required
fine-tuning.  A first conclusion that can be drawn is that there are
regions in which the fine-tuning is ameliorated with respect to the
MSSM, roughly reduced to twice the tuning of the NMSSM, $\Delta \sim
20-30$~\cite{Gherghetta:2012gb}.  Let us stress however that the
mechanism that allows for the increased naturalness is completely
different: while in the NMSSM it is due to the enhanced tree level
Higgs boson mass, here it is due to the collective loop enhancement
which reduces the sensitivity to the single mass involved. Moreover, 
we do not attempt to scan the parameter space to find the minimum 
achievable tuning of the model compatible with experimental data. 
It may well be that in some region of parameter space the tuning can be 
better that the one here presented. 
A further point which is worth mentioning is that stop masses in the TeV range
do not increase the fine-tuning, which is basically driven by
$m_{R_d}^2$, Eq.~(\ref{eq:FT_upperbounds}).  Together with the already
mentioned improved naturalness bound on the gluino mass, this makes
less worrisome the non observation so far of any superpartner at the
LHC.
 
What can we expect to observe at LHC-13, given this framework?  It is
of course quite difficult to make a robust a solid statement.  As we
have seen, since the fine-tuning is driven by $m_{R_d}$ and $m_{adj}$
in the interesting part of the parameter space, naturalness does not
require the stop to be as light as possible.  On the contrary, a
relatively heavy stop (with a mass around $1$ TeV) is preferred since
it can give a sizable contribution to the Higgs mass, allowing for the
state which are driving the fine-tuning to be lighter.  In any case,
we still expect $\mu$ to be as low as possible, with the Higgsino
possibly ``right around the corner''.

\acknowledgments We would like to thank Hugues Beauchesne, Marco
Farina, Tony Gherghetta, Yuri Shirman and Benedict Von Harling for
useful discussions.  E.B. acknowledges partial support by the Agence
National de la Recherche under contract ANR 2010 BLANC 0413 01 and by
the Spanish Ministry MICINN under contract FPA2010-17747.  T.G. is
supported in part by the Natural Sciences and Engineering Research
Council of Canada (NSERC).  This work was supported by the S\~ao Paulo
Research Foundation (FAPESP) under grant \#2011/11973 .  Fermilab is
operated by Fermi Research Alliance, LLC under Contract No.
DE-AC02-07CH11359 with the United States Department of Energy.

\appendix
\section{Potential minimization and mass matrices}\label{app:potential_mass}

We collect here useful formulas obtained from the minimization of the
tree level scalar potential.  For simplicity, we will take from the
beginning the limit $\tan\beta \gg 1$.

\vspace{2mm}
Using for the vacuum expectation values the convention $\langle h_u
\rangle = v$, $\langle t \rangle = v_T$, $\langle s \rangle = v_S$,
the minimization of the scalar potential,
Eq.~(\ref{eq:neutr_potential}) gives
\be\label{eq:minimization}
\begin{array}{ccl}
m_{H_u}^2 &=& \sqrt{2} \left( g M_{\tilde W} v_T -g' M_{\tilde B} v_S\right) - \frac{m_Z^2}{2} -\left( \lambda_S v_S + \lambda_T v_T + \mu \right)^2  \, , \\
v_T &=& \frac{\sqrt{2} g M_{\tilde W} -2 \lambda_S \lambda_T v_S   - 2 \lambda_T \mu }{2 \left(2 B_T+4 M_{\tilde W}^2 + m_T^2+ \lambda_T^2 v^2  \right)} v^2 \, ,\\
v_S &=& -\frac{\sqrt{2} g' M_{\tilde B} +2 \lambda_S \lambda_T v_T  + 2 \lambda_S \mu }{2 \left(2 B_S+4 M_{\tilde B}^2+m_S^2+\lambda_S^2 v^2 \right)} v^2 \, . 
\end{array}
\ee
The squared mass matrix for the CP-even scalars, in the $(h_u, t, s,
r_d)$ basis, reads
\be
M^2_{\rm CP-even}=
\left(
\begin{array}{cccc}
 m_Z^2 & \cdot & \cdot & \cdot \\
  v \left(2\lambda_T \left( \lambda_S v_S +\lambda_T v_T +\mu \right)-\sqrt{2}  g M_{\tilde W}\right) &
   m^2_{T_R}+\lambda_T^2 v^2  &
   \cdot  & \cdot \\
  v \left(2\lambda_S \left( \lambda_S v_S +\lambda_T v_T +  \mu \right)+ \sqrt{2} g' M_{\tilde B}+\right) & 
  \lambda_S \lambda_T v^2  & m^2_{S_R}+ \lambda_S^2 v^2 & \cdot \\
  0 & 0 & 0 & m^2_H
\end{array}
\right)\, ,
\ee
where $m_{T_R}^2$ and $m_{S_R}^2$ are defined below
Eq.~(\ref{eq:mh_approx}), while $m^2_H$, the mass of the CP-even inert
doublet, is given by
\be
\begin{array}{ccl}
m_H^2 &=& \mu^2 + m_{R_d}^2 -\frac{m_Z^2}{2} +\sqrt{2}\left( g M_{\tilde W} v_T - g' M_{\tilde B} v_S \right) 
+ 2 \left(\lambda_S v_S + \lambda_T v_T \right) \mu  + \\
&& \mbox{} + \left( \lambda_S v_S + \lambda_T v_T \right)^2 +  \left( \lambda_S^2 + \lambda_T^2 \right) v^2
\end{array}
\ee
Turning to the CP-odd squared mass matrix, in the $(a_t, a_s,
a_{r_d})$ basis it is
\be M^2_{\rm CP-odd}=
\left(
\begin{array}{ccc}
 m_T^2 - 2 B_T + \lambda_T^2 v^2 & \cdot & \cdot \\
 \lambda_S \lambda_T v^2 & m_S^2 - 2 B_S + \lambda_S^2 v^2 & \cdot \\
 0 & 0 & m_H^2
\end{array} \right) \, ,
\ee
with the CP-odd component of the inert doublet degenerate in mass with
the CP-even part.

\vspace{2mm} 
To conclude, the entries of the charged scalar squared mass matrix in
the basis $(H_u^+, T^+, (T^-)^*, (R_d^-)^*)$ are
\be
\begin{array}{ccl}
M^2_{11} &=& 2 v_T \left[ \sqrt{2} g M_{\tilde W} -2 \lambda_T \left( \lambda_S v_S +  \mu \right) \right] \\
M^2_{12} &=& -\frac{v}{2} \left[\sqrt{2} g^2 v_T- 2 g M_{\tilde W} +2\sqrt{2} \lambda_T \left( \lambda_S v_S - \lambda_T v_T + \sqrt{2} \mu \right) \right]  \\
M^2_{13} &=& \frac{v}{2} \left[\sqrt{2} g^2 v_T +2  g M_{\tilde W} -2\sqrt{2} \lambda_T \left( \lambda_S v_S + \lambda_T v_T + \sqrt{2} \mu \right) \right] \\
M^2_{22} &=& m_T^2 + 2 M_{\tilde W}^2+ 2 \lambda_T^2 v^2 +\frac{g^2}{2}\left( 2v_T^2-v^2 \right)  \\
M^2_{23} &=& 2\left( M_{\tilde W}^2 + B_T \right) - g^2 v^2 \\
M^2_{33} &=& m_T^2 + 2 M_{\tilde W}^2+ \frac{g^2}{2}\left( 2v_T^2+v^2 \right) \\
M^2_{44} &=& m_H^2 + m_W^2 -2 \sqrt{2} g M_{\tilde W} v_T - 4 \lambda_S \lambda_T v_S v_T - 4 \sqrt{2} \lambda_T \mu v_T + 
\left(\lambda_T^2 - \lambda_S^2 \right) v^2
\end{array}
\ee
with all the other entries vanishing.  The $3\times 3$ submatrix
obtained by taking out the $R_d^-$ entry has vanishing determinant as
expected, since one combination of the charged scalars is the would-be
Goldstone boson eaten up by the $W^\pm$.

\bibliography{Higgs}{}
\bibliographystyle{utphys_MOD}
\end{document}